# Obstructed surface states as the origin of catalytic activity in inorganic heterogeneous catalysts

Guowei Li[1,2,3,8], Yuanfeng Xu[4,5,8], Zhida Song[5], Qun Yang[1], Uttam Gupta[1], Vicky Süβ[1], Yan Sun[1], Paolo Sessi[4], Stuart S. P. Parkin[4], B. Andrei Bernevig[4,5,6,*], Claudia Felser[1*]

1. Max Planck Institute for Chemical Physics of Solids, Dresden, Germany
2. CAS Key Laboratory of Magnetic Materials and Devices, and Zhejiang Province Key Laboratory of Magnetic Materials and Application Technology, Ningbo Institute of Materials Technology and Engineering, Chinese Academy of Sciences, Ningbo 315201, China.
3. University of Chinese Academy of Sciences, Shijingshan District, Beijing 100049, China
4. Max Planck Institute of Microstructure Physics, Halle (Saale), Germany
5. Department of Physics, Princeton University, Princeton, NJ, USA
6. Donostia International Physics Center, P. Manuel de Lardizabal 4, 20018 Donostia-San Sebastian, Spain
7. IKERBASQUE, Basque Foundation for Science, Bilbao, Spain
8. These authors contributed equally to this work.

Email: Claudia.felser@cpfs.mpg.de, bernevig@princeton.edu

**Abstract:**

**The discovery of new catalysts that are efficient, sustainable, and low cost is a major research endeavor for many industrial chemical processes[1,2]. This requires an understanding and determination of the catalytic origins for the given catalysts, which still remains a challenge[3,4]. Here we describe a novel method to identify new catalysts based on searching for crystalline symmetry-protected obstructed atomic insulators (OAIs) that have metallic surface states on otherwise semiconducting or insulating compounds[5–9]. The Wannier charge centers in OAIs are pinned by symmetries at some empty Wyckoff positions so that surfaces that accommodate these sites are guaranteed to have metallic obstructed surface states (OSSs)[10].**

**Beyond the well-studied 2H-MoS$_2$[10], we further verified our theory on the catalysts, 2H-MoTe$_2$, and 1T′-MoTe$_2$, whose catalytic active sites are consistent with our calculations of obstructed Wannier charge centers (OWCCs) and OSSs. In addition, we have predicted the location of catalytic active sites and confirmed these predictions by exploring the hydrogen evolution reaction on NiPS$_3$ bulk single crystals, which we find to be one of the most promising new catalysts with high activity and, moreover, of low cost. Most importantly, we successfully identified several high-efficient catalysts just by considering the number of OWCCs and the crystal symmetry of the OAIs. Using the real space invariant (RSI) theory and high-throughput computational methods applied to a database of 34013 topologically trivial insulators, we have identified 1788 unique OAIs (3383 ICSD entries)[10], of which 465 are potential high-quality catalysts for heterogeneous reactions. The miller indices of the active surfaces are also obtained. Our new methodology will facilitate and accelerate the discovery of new catalysts for a wide range of heterogeneous redox reactions, where sustainability, toxicity, and cost must be considered.**

High-performance heterogeneous catalysts are key to the preparation of many chemicals that are the foundation of, amongst others, photo- and electro- chemical water splitting, fuel cells, hydrogenation, and the Haber-Bosch process[2,11,12]. A prerequisite for the design of highly active catalysts is the identification of the origin of the activity. However, this remains a challenge[13,14]. The activity of a given catalyst is traditionally associated with the properties of its surfaces. Thus, materials with large surface areas, good conductivity, and high mobility are understood to be good catalysts, as they have abundant active sites that favor the adsorption of intermediates and electron transfer in redox reactions. This is the motivation for widely used catalyst synthesis strategies such as nano-structuring, doping, alloying, or adding defects. Each method aims to either expose preferential crystal surfaces or to engineer them to make them more active[15,16]. However, it is still a formidable task to locate the position of active sites rapidly and precisely from the design perspective, making the discovery of high-performance catalysts from the many potentially interesting materials a challenge.

Topological materials have robust surface states and massless electrons with high mobilities[17]. Moreover, many state-of-the-art catalysts (such as Pt, Pd, Cu, Au, $IrO_2$, and $RuO_2$) are understood, either from theory or experiment, to have topologically derived surface states (TSSs)[18,19]. Thus, there is some evidence for the important role of TSSs in catalytic reactions[20]. Such states are mainly composed of highly delocalized *sp*-orbitals (noble metals) or *d* orbitals with strong spin-orbit coupling (SOC) (noble metal oxides). Furthermore, these states are fundamentally derived from the band structure of the bulk material, and, thus, must be distinguished from what one might term "topologically trivial" surface modifications that form the basis of many of the common strategies to increase catalytic activity, mentioned above[21,22].

In this work, we propose that the OSSs of OAIs form a new class of active sites for inorganic heterogeneous catalysts. Similar to TSSs, the OSSs originate from the topologies of bulk electronic bands and provide open sites for molecular adsorption. The OSSs can exist in clean, large-gap, topologically trivial insulators and are separated from the bulk states in the energy spectrum, unlike TSSs, which are always connected to bulk states with a small bandgap. Using the Topological Quantum Chemistry (TQC) theory and the RSIs developed in our previous works [5,6], we have identified 1788 unique OAIs (3383 ICSD entries), of which 465 unique compounds have promising catalytic activities. Most importantly, we also predict their Miller indices of specific cleaved surfaces which have active sites. Thus, our work gives a clear direction for both the understanding and the design of high-performance catalysts from the numerous known inorganic compounds.

We address this problem by taking both the bulk electronic structure and the crystal geometry into consideration. As schematically shown in Figure 1A, for all the compounds with crystal structures provided in the Inorganic Crystal Structure Database (FIZ Karlsruhe, Germany)[23], we have calculated their electronic structures and symmetry eigenvalues with the data now evaluable to the public via the TQC database (www.topologicalquantumchemistry.com)[6], where all the materials are classified into trivial and topological materials (insulators, semimetals, metals). In the next step, for all the band representations of topologically trivial insulators, we calculate their 3D RSIs[5,6,10], which characterize the multiplicities of symmetric Wannier functions pinned at the real-space positions that are referred to as the obstructed Wannier charge centers (OWCCs)[10]. Finally, we have provided the Miller indices of cleavage planes that cut through the OWCCs and lead to the metallic OSSs.

The usual dangling bond can be thought of as the simplest case of OSSs, where the OWCC resides between two bonding atoms and the OSSs at the surface are just the broken bonds and will be localized around the bonding atoms. However, the general structures of OSSs can be exotic and very different from simple dangling bonds. OSSs are distinguished from TSSs by the fact that their surface states do not fill the entire energy gap between the conduction and the valence band (Figure 1B). Thus, the crystal surface with OSSs is characterized by high conductivity, open sites for bonding and adsorption, and a high density of electronic states around the Fermi level. Therefore, the OSSs provide a straightforward path to determine the position of active sites and to guide the design of highly active catalysts by choosing desired crystal surfaces (Figure 1C). Using the RSI indices which, given a band structure, immediately determine the position of the OWCCs, we have discovered 1788 unique OAIs (3383 ICSDs)[10], and 465 of them as potential catalysts with well-defined catalytic active sites. In this letter, several representative single crystals, OAI 2H-MoTe$_2$, topological semimetal 1T′-MoTe$_2$, and OAI NiPS$_3$ are chosen for the experimental validation of the hypothesis. Most importantly, high-performance catalysts are predicted successfully by condensing the location and density of OSSs, including the OAIs RuP$_4$, and FeP$_4$. This work provides a descriptor to find the active sites of a given catalyst rapidly, making the design of highly active catalysts more efficient.

**Results**

**OSSs in two-dimensional 2H-MoTe$_2$**

The van der Waals compounds including 2H-MoS$_2$, 2H-WS$_2$, and 2H-MoTe$_2$ [ICSD 105091, SG (Space Group) 194 (*P*63/*mmc*)] have been the subject of extensive research studies because of their high intrinsic catalytic efficiency for industrial-scale catalysis reactions such as hydro-desulfurization and HER[24–27]. Electrocatalytic activity measurements on various 2H phase nanostructured catalysts have confirmed that only the edge surfaces are catalytically active towards HER, while the thermodynamically stable (001) surface has no such activity. Only when the basal surfaces are modified by defect engineering such as by making vacancies, can they be activated and optimized for catalysis because of the appearance of surface states[28]. In Figure 2A, the crystal structure of 2H-MoTe$_2$ and the position of OWCCs are displayed. The center of charge is localized at the *2b* position, that has no atom occupation (Table S1-2). This is confirmed by our surface states calculation on the (100) edge surface and (001) basal surface, as shown in Figure 2B and C. The OSSs only exist at the (100) edge surface and are located close to the Fermi level. As an HER catalyst, we predict that only the edge surfaces of the 2H-MoTe$_2$ are catalytically active for hydrogen production. To confirm this prediction, we synthesized high-quality MoTe$_2$ bulk crystals in a pure 2H phase[29]. The single crystal was attached to a Ti wire and served as the working electrode. To obtain the HER current contribution from the edge surfaces and basal surfaces, the basal plane and edges of the single crystals were covered by gel, respectively. Linear sweep voltammetry (LSV) curves recorded using a 0.5 M H$_2$SO$_4$ electrolyte indicate that the activity of the edge surfaces is more or less the same as the whole crystal, and is much larger than that for the (001) basal surface (Figure 2B). A photo of the crystal was taken during the chronopotentiometry measurements. It can be seen clearly that the hydrogen bubbles are produced only at the edge surfaces (Figure 2C). When the large basal plane was covered by gel, one can still see the hydrogen evolution at the edge surfaces (Figure S1), which is consistent with the LSV results. Electrochemical impedance measurements were carried out on the whole crystal, the edge surfaces, and the basal surface. The corresponding Nyquist plots suggest that the edge surface has a much smaller charge transfer resistance than the basal surface, indicating a much higher conductivity at the edge surfaces (Figure S2). The experimentally observed catalytic behavior of 2H-MoTe$_2$ is in perfect agreement with our theoretical concept. Also for the isostructural compound 2H-MoS$_2$, our theory suggests that the OWCCs and OSSs are located at the edge surfaces, as we observe for 2H-MoTe$_2$.

**The prediction of catalytic active surfaces**

With this principle in mind, it is possible to determine the catalytic active surfaces of a given catalyst quickly. NiPS$_3$ crystallizes in the symmetries of the SG 12 (*C2/m*). Applying the RSI theory in Tables S3 and S4, we find that all the atoms have different intercepts with the 2a position in the (001) direction, and only the *2a* position has a nonzero RSI and is not occupied by atoms. Thus, side surfaces that are perpendicular to the (001) surface should be equipped with the OSSs and be catalytically active towards the HER. Indeed, LSV curves vividly confirmed that the HER activities originate from the edge surfaces with the observation of hydrogen bubbles only at the predicted positions (Figure 3A)[30]. In agreement with the theoretical predictions (Table S3-4), we observed an increased conductivity at selected edges on the exfoliated $NiPS_3$ flake using conductive Atomic Force Microscopy (AFM), supporting the existence of metallic electronic states (Figure 3B, Figure S3).

Most van der Waals compounds expose their thermodynamically stable basal planes, which have no charge density for bonding and charge transfer with adsorbed molecules and are thus catalytically inert according to our theory. Based on our RSI calculations of the OAIs, we predict, however, that there is a class of van der Waals compounds that have surface states in their basal planes. The topological Weyl semi-metal, the 1T′ phase of $MoX_2$ (X = S, Se, Te), which is obtained by breaking the inversion structural symmetry of $2H-MoTe_2$, is a good candidate for testing our hypothesis[31]. In both polytypes, the Mo atoms are located at the center of a framework defined by two Te atom triangles to form an $[MoTe_6]$ octahedral unit. In the $2H-MoTe_2$ phase, the $[MoTe_6]$ units have a trigonal prismatic structure whereas in the 1T′ phase these units form distorted octahedra. This results in a charge density and metallic surface states on both (001) and (100) surfaces in the latter 1T′ structure (Figure S4)[26,32]. We synthesized high-quality $1T′-MoTe_2$ single-crystals with megascopic (001) basal plane that we sconfirmed with Raman spectroscopy (Figure S5). Indeed, HER measurements on the edges and basal planes of a bulk $1T′-MoTe_2$ single-crystal confirm the prediction of our OSS theory that all the crystal surfaces are active for hydrogen production (Figure 3C). In addition, we observed a significantly lower charge transfer resistance for the (001) plane than the edge surfaces, suggesting a high conductivity at the basal plane (Figure 3D).

**Searching for highly active catalysts with OSSs**

New catalysts with high efficiencies can now be quickly predicted by identifying the position of the OSSs. The objective is to identify materials that have high numbers of OWCCs on a given crystal facet plus as many of these facets, as possible. The family of $3d$ and $4d$ transition metal tetra-phosphides MP$_4$ (M = Fe, Ru, Mn…) is a promising choice to meet this objective, and some of them contain naturally abundant and, therefore, low-cost elements. Our theory suggests that the surfaces that cut the $2d$ position but without overlapping with the atoms have surface states (Tables S5 and S6). This indicates that the OWCCs are located between the P-P bonds (Figure 4A), which means that all crystal surfaces with broken P-P bonds possess OSSs and should be catalytically active. This includes, but is not limited to the surfaces, (001), (010), and (011). We grew nanostructured RuP$_4$ on a Ni foam to increase the density of active sites (Figure S6-7). The overpotentials required to produce cathodic current densities of 10 mA cm$^{-2}$ are 28 and 31 mV using 0.5 M H$_2$SO$_4$ and 1 M KOH electrolytes, respectively, which are comparable to one of the best performing catalysts, Pt/C (Figure 4B, Figure S8). The turnover frequency (TOF) of the HER using RuP$_4$ as the catalyst was determined by normalizing the kinetic current to the electrochemically active surface area (Figure S9-10). At an overpotential of 100 mV, the TOFs were measured to be 1.86 and 7.8 H$_2$ s$^{-1}$ in acidic and alkaline electrolytes, respectively. These values are comparable to those reported for state-of-the-art catalysts such as Pt/C (Figure 4C).[33–37] In addition, the RuP$_4$ nanostructures exhibited good electrochemical stability as an HER catalyst, without significant activity loss during a 25 h measurement (Figure S11).

Having identified RuP$_4$ as a high-performing catalyst we considered other members of this family containing only highly abundant elements. Moreover, we identified FeP$_4$ as a highly interesting candidate for photocatalysis water splitting, since it has a suitable bandgap of ~ 1.8 eV, matching the spectra of visible light. We prepared a series of photocatalysts including the OAIs, $FeP_4$, $MoS_2$, and $FeS_2$ (Table S7-8) as well as the trivial (non-obstructed) insulators $Fe_2O_3$ and FeS (Figure S12). Typical temporal evolution of $H_2$ curves suggests an impressive activity of the synthesized $FeP_4$ microcrystals (Figure S13). When scaled by the surface area (Figure S14), all the OAIs exhibit dramatically better intrinsic activities than normal insulators, as shown in Figure 4D.

**Conclusion and outlook**

We have demonstrated our concept of identifying novel catalysts using calculations of OSSs with just a few highly promising examples that span stoichiometric compounds with crystal structures ranging from 2D van der Waals to phosphides. Beyond these examples, we have identified 465 OAIs with promising catalytic activities from among the known 34013 topologically trivial insulators. The positions of the OSSs for these 465 OAIs are listed in Appendix A. And most excitingly, a subset of these selected compounds, including the iron-sulfur minerals FeS and $FeS_2$, exhibit chiral OSSs due to the loss of mirror symmetries at some of their surfaces. Though the compounds have achiral cubic space groups, one can speculate that OSSs on the surfaces of the chiral plane group can be used for the synthesis of chiral molecules[37,38]. Our work demonstrates that metallic OSSs derived from bulk symmetry calculations are the catalytic activity origin of inorganic heterogeneous catalysts. We conjecture that OSSs can be used for other electrocatalytic reactions beyond HER.

**Acknowledgements**

This work was financially supported by the European Research Council (ERC Advanced Grant No. 742068 'TOPMAT'). We also acknowledge funding by the DFG through SFB 1143 (project ID. 247310070) and the Würzburg-Dresden Cluster of Excellence on Complexity and Topology in Quantum Matter ct.qmat (EXC2147, project ID. 39085490) and via DFG project HE 3543/35–1. BAB's work is part of a project that has received funding from the European Research Council (ERC) under the European Union's Horizon 2020 research and innovation program (grant agreement n° 101020833). UG thanks Dr. Guangbo Chen and Prof. Xinliang Feng for measuring the specific surface areas.

**Author contributions**

C.F. and B.A.B. conceived this work; G.L. designed and performed the experimental works. Y.X. carried out the theoretical calculations and wrote the theory part along with B.A.B. and Z.S. V. S. provided the 2H and 1T′-MoTe$_2$ single crystals. U.G. contributed photocatalysis results. P.S. and S.S.P.P. performed the STM and AFM measurements on a NiPS$_3$ crystal. G. L. and Y.X. co-wrote the manuscript with C.F. and S.S.P.P.. All authors discussed the results and contributed to the final manuscript.

**Funding**

Open access funding provided by Max Planck Society.

**Competing interests**

The authors declare no competing interests.

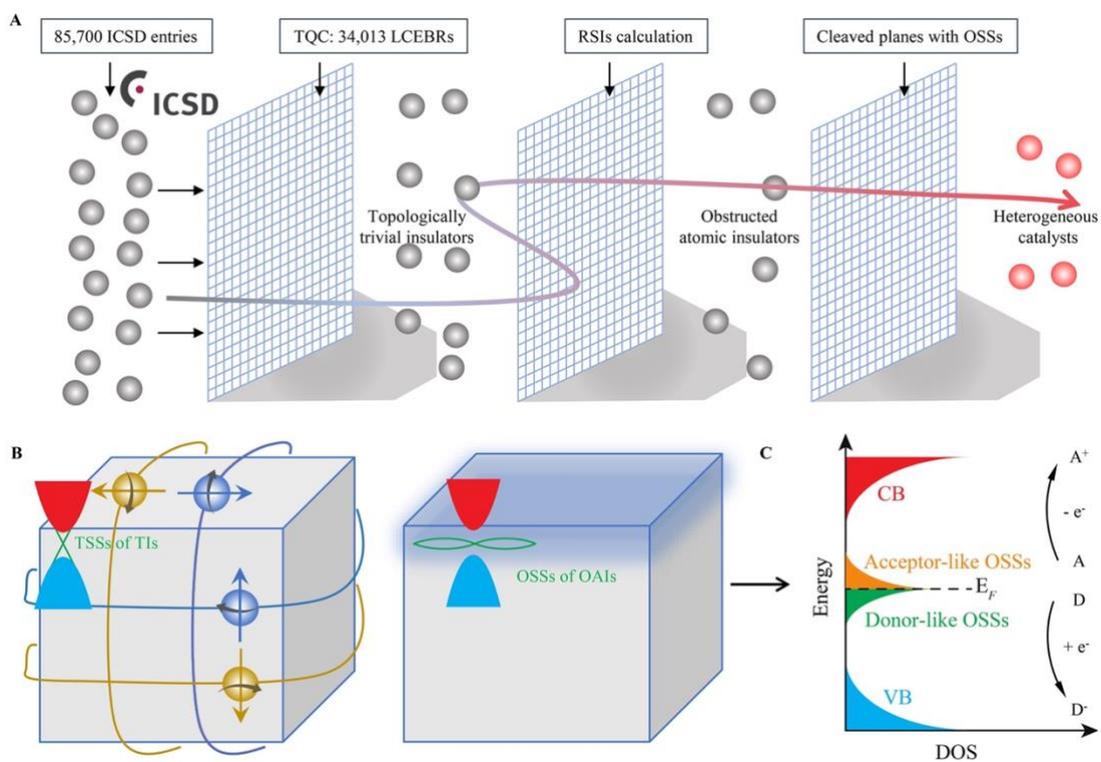

**Figure 1. The design of heterogeneous catalysts from obstructed atomic insulators (OAIs).** (A) Screening of potential OAIs and heterogeneous catalysts from the Topological quantum chemistry website (www.topologicalquantumchemistry.fr) by calculating the RSIs. (B) Comparisons between the TSSs and OSSs. (C) Illustration of the role of OSSs in OAIs for reduction and oxidation catalytic reactions.

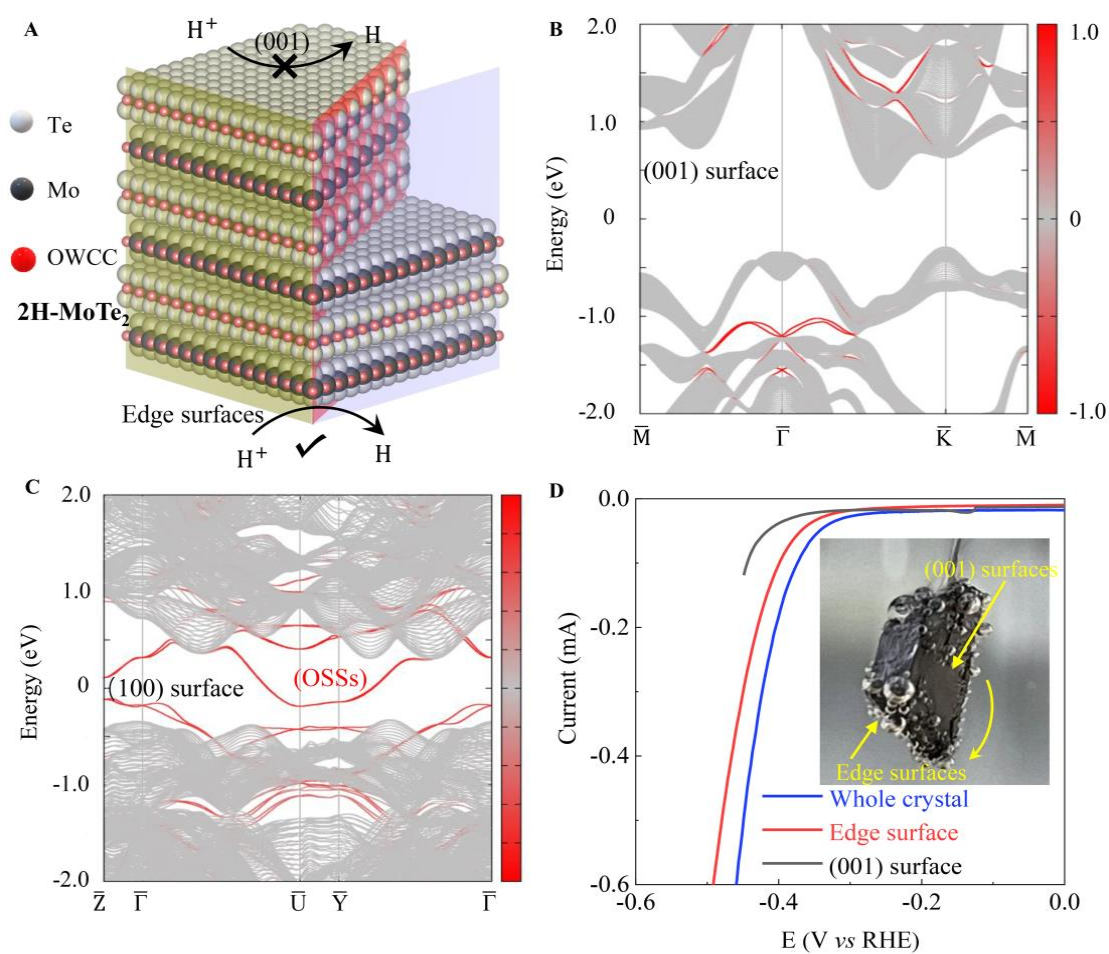

**Figure 2. The relationship between OWCCs and active sites in 2H-MoTe$_2$ for hydrogen evolution.** (A) The crystal structure of 2H-MoTe$_2$ and the position of OWCCs. As an HER catalyst, the (001) basal plane is catalytically inert. All the edge surfaces are active toward hydrogen evolution. The obstructed surface states calculation of 2H-MoTe$_2$ at the (B) (001) and (C) (100) surfaces, respectively. The OSSs only exist at the edges surfaces, such as the (100) surface. The gray and red lines represent the respective bulk and surface bands. (D) LSVs of the 2H-MoTe$_2$ HER catalyst. It can be seen clearly that almost all the activity originates from the edge surfaces. Photo of the 2H-MoTe$_2$ bulk single crystal during the hydrogen evolution process (with a constant overpotential of -0.43 V *vs*. RHE) is shown as an inset. Hydrogen bubbles are observed at the edge surfaces, while the basal (001) surface is not active for hydrogen production.

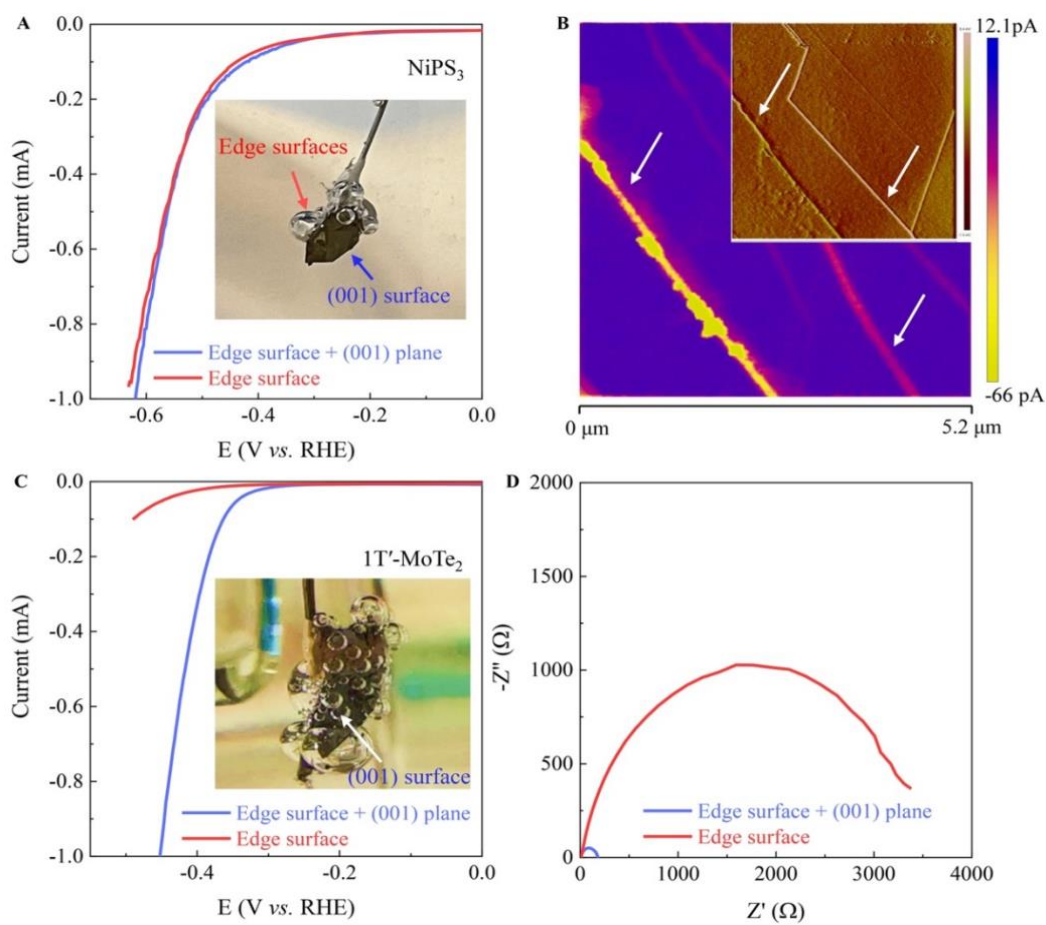

**Figure 3. The prediction and experimental verification of catalytic active sites.** (A) Polarization curves of the NiPS$_3$ bulk crystal were recorded from the edge surfaces and the whole crystal (edge surfaces plus (001) surfaces). The inset shows that the HER activities are confined predominantly to the edge surfaces, although the areas of the (001) basal planes are much bigger than those of the side surfaces. (B) The conductive AFM image of the exfoliated NiPS$_3$ crystal surface. The conductivity at the steps is significantly higher than the (001) basal plane. The inset figure shows the corresponding AFM image. (C) Polarization curves of 1T′-MoTe$_2$ bulk crystal for the whole crystal and the edge surfaces. It can be clearly seen that the HER activities are contributed almost entirely by the (001) basal plane. The inset confirms that the hydrogen bubbles are produced at the (001) plane. (D) Corresponding impedance spectra of the 1T′-MoTe$_2$ bulk crystal recorded at different surfaces. One can see a much smaller charge transfer resistance when exposing the metallic (001) surface.

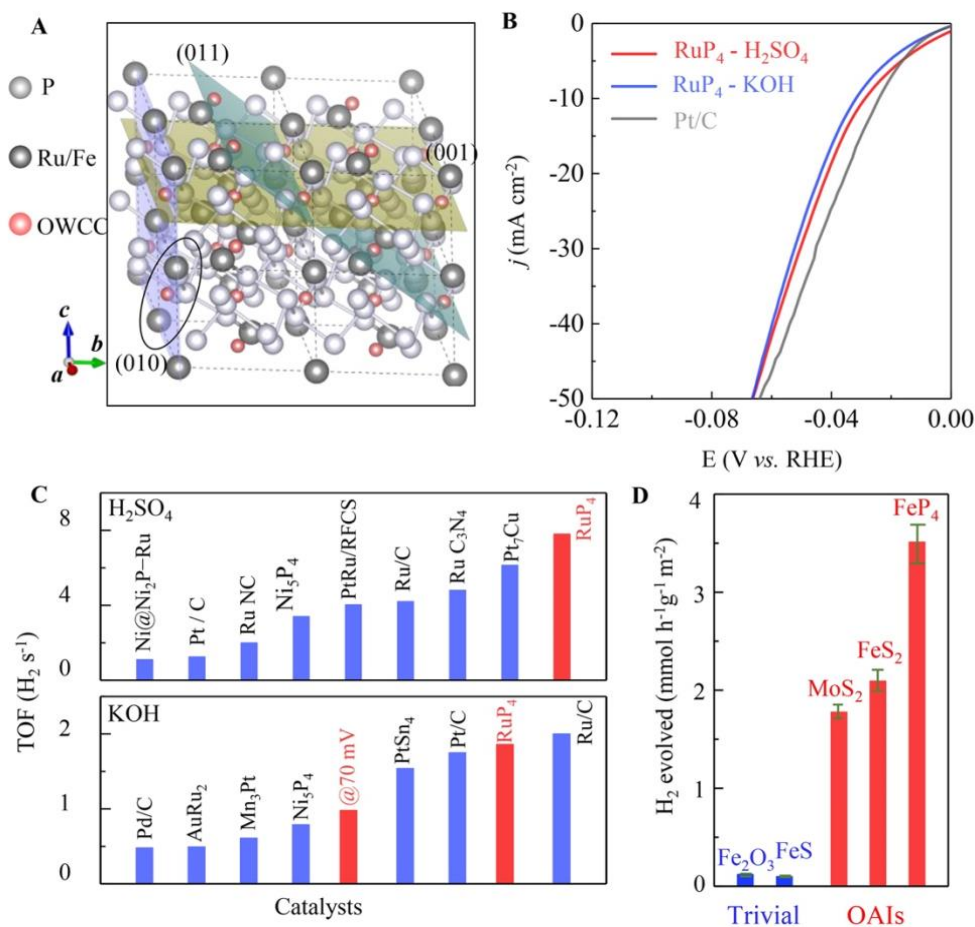

**Figure 4. Prediction of high-performance HER catalysts from calculations of OSSs.** (A) The position of OWCCs and cleavage planes giving rise to OSSs in FeP$_4$ and RuP$_4$. (B) Comparison of LSV curves for Pt/C and RuP$_4$ nanostructures in acid and alkaline conditions. (C) Comparison of the TOF values for RuP$_4$ and state-of-the-art catalysts. (D) Relative HER efficiency of the OAIs FeP$_4$, MoS$_2$, and FeS$_2$, and the trivial insulators Fe$_2$O$_3$ and FeS (scaled by specific surface areas).

# Supplementary materials for "Obstructed surface states as the origin of catalytic activity in inorganic heterogeneous catalysts"

**Methods**

**Synthesis of NiPS$_3$ single crystals**

Stoichiometric amounts of Ni, P, and S powders were mixed and sealed in a quartz tube under a high vacuum. The tube was placed in a tube furnace and heated to 700 °C in 40 minutes and then kept at this temperature for 7 days. Plate-like single crystals can be obtained when the tube is cooled down to room temperature.

**Synthesis of 1T′-MoTe$_2$ single crystals**

1T′-MoTe$_2$ crystals were grown via the chemical vapor transport method. polycrystalline MoTe$_2$ powder was first synthesized by heating stoichiometric amounts of Mo and Te pieces in an evacuated quartz tube at 800 °C for 7 days. For the growth of single crystals, MoTe$_2$ powder and TeCl$_4$ were mixed in a quartz ampoule and heated in a two-zone furnace. Crystallization was conducted from (T2) 1,000 to (T1) 900 °C. The quartz ampoule was then quenched in ice water to yield the high-temperature monoclinic phase.

**Synthesis of 2H-MoTe$_2$ single crystals**

2H-MoTe$_2$ crystals were grown using a similar method to that of 1T′-MoTe$_2$ but without quenching.

**Synthesis of RuP$_4$ polycrystalline powder**

0.2 g of RuCl$_3$, 1.5 g (NH$_4$)$_2$HPO$_4$, and 0.1 g melamine were dissolved in distilled water with a Ni form. The solution was dried at 60 °C and then the powders were transferred to a tube furnace. The tube was heated to 500 °C in 40 minutes and then kept at this temperature for 2 h with a hydrogen flow of 60 mL/minute. The Ni form and powders are washed with water and ethanol several times after cooled down to room temperature.

**Synthesis of FeP$_4$ polycrystalline powder**

Powders of Fe (Alfa Aesar purity >99%) and P (Alfa Aesar purity > 99%) in stoichiometric amounts and a trace amount of I$_2$ (Alfa Aesar purity > 99%) were mixed in a quartz ampoule and heated from 700-600 °C in a two-zone furnace for 1 week. The compound was characterized using powder X-ray Diffraction.

**Synthesis of Fe$_2$O$_3$ polycrystalline powder**

Fe$_2$O$_3$ (assay >96%) was commercially procured from Kali Chemie. The powder was heated at 900 °C for two hours at rate of 100 °C/h and cooled to room temperature at the same rate. The compound was characterized using powder X-ray Diffraction.

**Synthesis of MoS$_2$ polycrystalline powder**

MoS$_2$ powder was synthesized by heating mixture of elemental Mo (Alfa Aesar purity >99%), S (Alfa Aesar purity >99%) and trace amount of I$_2$ (Alfa Aesar purity >99%) for 2 weeks from 860-725 °C in a two-zone furnace. The compound was characterized using powder X-ray Diffraction.

**Synthesis of FeS$_2$ polycrystalline powder**

Elemental of Fe (Alfa Aesar purity >99%) and S (Alfa Aesar purity >99%) in stoichiometric ratio along with trace amount of I$_2$ and heated at 600 °C for 120 hours. The compound was characterized using powder X-ray Diffraction.

**Synthesis of FeS polycrystalline powder**

Elemental Fe and S in stoichiometric ratio along with trace amount of I$_2$ and heated at 950 °C for 120 hours. The compound was characterized using powder X-ray Diffraction.

**Electrochemical catalytic measurements**

All the electrochemical HER catalytic measurements were performed on an Autolab PGSTAT302N electrochemistry workstation an Ar saturated 0.5 M H$_2$SO$_4$ or 1 M KOH solution. An Ag/AgCl (3 M KCl) electrode and graphite rod as the reference electrode and counter electrode, respectively. For the testing of bulk single crystals, the crystals are attached to a Ti wire with silver paint and served as the working electrode. Linear sweep voltammograms were recorded with a scan rate of 1 mV / S. The electrochemical impedance spectroscopy (EIS) measurements were conducted from 100 kHz to 0.1 Hz. The amplitude of the sinusoidal potential signal was 10 mV. All potentials were referenced to a reversible hydrogen electrode according to E (versus RHE) = E (versus Ag/AgCl) + (0.210 + 0.059 pH) V.

**Estimation Turnover frequency (TOF)**

For RuP$_4$, we assumed all P sites are active towards HER. The double-layer capacitance (C$_{dl}$) was extracted from CV scanning in the potential range with non-faradaic processes happened at various scan rates. The specific capacitance can be converted into an electrochemical active surface area (ECSA) using the specific capacitance value for a flat standard with 1 cm$^2$ of real surface area.

The total number of hydrogens turn-overs was calculated from the current density according to:

$$\# H_2 = \left(j\ \frac{mA}{cm^2}\right)\left(\frac{1\ Cs^{-1}}{1000\ mA}\right)\left(\frac{1\ mol\ e^-}{96485.3\ C}\right)\left(\frac{1\ mol\ H_2}{1\ mol\ e^-}\right)\left(\frac{6.022\times 10^{23}\ H_2}{1\ mol\ H_2}\right)$$

$$= 3.12\times 10^{15}\ \frac{H_2/s}{cm^2} per\ \frac{mA}{cm^2}$$

Then the HER turnover frequency (TOF) as a function of current density is defined as:

$$TOF = \frac{\left(3.12\times 10^{15}\ \frac{H_2/s}{cm^2} per\ \frac{mA}{cm^2}\right)\times |j|}{\#\ active\ sites\ \times A_{ECSA}}$$

**Photocatalytic reaction measurements:**

In typical photochemical hydrogen evolution, 4 mg of the sample was taken in a 10 ml solution of triethanolamine (15% v/v) and was sensitized by Eosin Y dye (14µM). The mixture was illuminated with 100W (working at 50% power) white LED (power density of 150 mW/cm$^2$) at constant stirring conditions. The evolved hydrogen gas was measured at a regular interval of one hour by manually injecting evolved hydrogen gas in Perkin Elmer gas chromatograph (580 GC). The activity of the catalyst was measured from the hydrogen evolution curve with time.

$$\text{Activity (mmol } g^{-1} h^{-1}) = \frac{\text{Hydrogen gas evolved per hour}}{\text{Mass of the catalyst used}}$$

| Multiplicity | Wyckoff letter | Site symmetry | Coordinates |
|---|---|---|---|
| 24 | l | 1 | (x,y,z) (-y,x-y,z) (-x+y,-x,z) (-x,-y,z+1/2) (y,-x+y,z+1/2) (x-y,x,z+1/2) (y,x,-z) (x-y,-y,-z) (-x,-x+y,-z) (-y,-x,-z+1/2) (-x+y,y,-z+1/2) (x,x-y,-z+1/2) (-x,-y,-z) (y,-x+y,-z) (x-y,x,-z) (x,y,-z+1/2) (-y,x-y,-z+1/2) (-x+y,-x,-z+1/2) (-y,-x,z) (-x+y,y,z) (x,x-y,z) (y,x,z+1/2) (x-y,-y,z+1/2) (-x,-x+y,z+1/2) |
| 12 | k | .m. | (x,2x,z) (-2x,-x,z) (x,-x,z) (-x,-2x,z+1/2) (2x,x,z+1/2) (-x,x,z+1/2) (2x,x,-z) (-x,-2x,-z) (-x,x,-z) (-2x,-x,-z+1/2) (x,2x,-z+1/2) (x,-x,-z+1/2) |
| 12 | j | m.. | (x,y,1/4) (-y,x-y,1/4) (-x+y,-x,1/4) (-x,-y,3/4) (y,-x+y,3/4) (x-y,x,3/4) (y,x,3/4) (x-y,-y,3/4) (-x,-x+y,3/4) (-y,-x,1/4) (-x+y,y,1/4) (x,x-y,1/4) |
| 12 | i | .2. | (x,0,0) (0,x,0) (-x,-x,0) (-x,0,1/2) (0,-x,1/2) (x,x,1/2) (-x,0,0) (0,-x,0) (x,x,0) (x,0,1/2) (0,x,1/2) (-x,-x,1/2) |
| 6 | h | mm2 | (x,2x,1/4) (-2x,-x,1/4) (x,-x,1/4) (-x,-2x,3/4) (2x,x,3/4) (-x,x,3/4) |
| 6 | g | .2/m. | (1/2,0,0) (0,1/2,0) (1/2,1/2,0) (1/2,0,1/2) (0,1/2,1/2) (1/2,1/2,1/2) |
| 4 | f | 3m. | (1/3,2/3,z) (2/3,1/3,z+1/2) (2/3,1/3,-z) (1/3,2/3,-z+1/2) |
| 4 | e | 3m. | (0,0,z) (0,0,z+1/2) (0,0,-z) (0,0,-z+1/2) |
| 2 | d | -6m2 | (1/3,2/3,3/4) (2/3,1/3,1/4) |
| 2 | c | -6m2 | (1/3,2/3,1/4) (2/3,1/3,3/4) |
| 2 | b | -6m2 | (0,0,1/4) (0,0,3/4) |
| 2 | a | -3m. | (0,0,0) (0,0,1/2) |

**Table S1.** The Wyckoff positions and their site symmetry groups of SG 194 (*P*6$_3$/*mmc*).

| Index | Wyckoff positions | Irreps and multiplicity | | | | | | | | | | | | | | | | mod |
|---|---|---|---|---|---|---|---|---|---|---|---|---|---|---|---|---|---|---|
| | | $\bar{\Gamma}_7$ | $\bar{\Gamma}_8$ | $\bar{\Gamma}_9$ | $\bar{\Gamma}_{10}$ | $\bar{\Gamma}_{11}$ | $\bar{\Gamma}_{12}$ | $\bar{A}_4\bar{A}_5$ | $\bar{A}_6$ | $\bar{H}_4\bar{H}_5$ | $\bar{H}_6\bar{H}_7$ | $\bar{H}_8$ | $\bar{H}_9$ | $\bar{K}_7$ | $\bar{K}_8$ | $\bar{K}_9$ | $\bar{L}_3\bar{L}_4$ | $\bar{M}_5$ | $\bar{M}_6$ | |
| $\delta_1$ | a | 0 | 0 | 0 | -1/2 | 1/2 | 1/2 | 0 | -1 | 0 | 0 | 0 | 0 | 0 | 0 | 0 | 0 | 0 | 1/2 | |
| $\delta_2$ | a | -1 | 0 | 0 | -1/4 | -1/4 | -1/4 | 1 | 0 | 0 | 0 | 0 | 0 | 0 | 0 | 0 | 0 | 0 | 1/4 | |
| $\delta_3$ | b | 1/3 | 0 | 0 | 1/3 | 1/3 | -1/3 | 0 | 2/3 | -1/3 | -1/3 | 0 | 0 | 0 | -2/3 | 0 | 0 | 0 | 0 | |
| $\delta_4$ | c | -2/3 | 0 | 0 | -2/3 | 1/3 | -1/3 | 1 | -1/3 | 2/3 | -1/3 | 0 | 0 | 0 | 1/3 | 0 | 0 | 0 | 0 | |
| $\delta_5$ | d | -2/3 | 0 | 0 | -2/3 | 1/3 | -1/3 | 1 | -1/3 | -1/3 | 2/3 | 0 | 0 | 0 | 1/3 | 0 | 0 | 0 | 0 | |
| $\delta_6$ | e | -2/3 | 0 | 0 | 1/3 | -2/3 | -1/3 | 0 | 2/3 | -1/3 | -1/3 | 0 | 0 | 0 | 1/3 | 0 | 0 | 0 | 0 | |
| $\delta_7$ | f | -1/3 | 0 | 0 | -1/3 | -1/3 | 1/3 | 0 | 1/3 | 1/3 | 1/3 | 0 | 0 | 0 | -1/3 | 0 | 0 | 0 | 0 | |
| $\delta_8$ | g | 0 | 0 | 0 | 1/4 | 1/4 | 1/4 | 0 | 0 | 0 | 0 | 0 | 0 | 0 | 0 | 0 | 0 | 0 | -1/4 | |
| $\eta_1$ | h | 1/3 | 0 | 0 | 4/3 | 4/3 | 2/3 | 0 | 2/3 | 2/3 | 2/3 | 0 | 0 | 0 | 4/3 | 0 | 0 | 0 | 0 | mod 2 |
| $\zeta_1$ | h,i,j,k | 7/3 | 0 | 0 | 4/3 | 4/3 | 8/3 | 0 | 8/3 | 8/3 | 8/3 | 0 | 0 | 0 | 4/3 | 0 | 0 | 0 | 0 | mod 4 |

**Table S2.** The formula of the RSIs of SG 194 (P6$_3$/mmc).

2H-MoTe$_2$ has the symmetries of the SG 194 ($P6_3/mmc$). As shown in Table S1, the Mo and S atoms occupy the 2c and 4f Wyckoff postions, respectively. The band structure and the band representation analysis of this material, which show it is a topologically trivial insulator with an indirect gap ~0.66 eV, can be found on www.topologicalquantumchemistry.com. We use the symmetry-data-vector to represent the irreps formed by the occupied Bloch bands at high symmetry momenta. In the orders of the irreps listed in Table S2, the symmetry-data-vector is

$$B = \left(m(\bar{\Gamma}_7), m(\bar{\Gamma}_8), m(\bar{\Gamma}_9), m(\bar{\Gamma}_{11}), m(\bar{\Gamma}_{12}) \cdots m(\bar{M}_5), m(\bar{M}_6)\right)^T$$
$$= (2,3,4,2,4,3,2,7,3,3,6,6,6,5,7,9,9,9)^T$$

where $m(\rho)$ represents the multiplicity of the irrep $\rho$ formed by the occupied Bloch bands at the corresponding high symmetry momentum. The expressions to calculate RSIs at all Wyckoff positions are defined in Fig. 1A. Nonzero RSIs at a Wyckoff position implies Wannier functions pinned at this position. (See Ref. [Xu, Y. et al. Three-dimensional real space invariants, obstructed atomic insulators and a new catalytic principle. Submitted (2021)] for the rigorous definition of RSIs.) Applying the RSI formulae in Table S2, we find that only the 2b position has a nonzero RSI, $\delta_3 = 1$. Since the 2b position is not occupied by atoms, the material 2H-MoTe$_2$ is in the OAI phase.

Now we analyze on which surfaces one can observe the OSSs. We first consider the [001] direction. The 2b position projects to {1/4, 3/4} (mod 1) in the (001) direction. Thus one may want to cleave the material at these intercepts to create a surface with OSSs. However, the Mo atoms occupying the 2c position also project to {1/4, 3/4} in the (001) direction. Thus such cleavages are unrealistic since one cannot cut an atom into two. On the other hand, there are OSSs on some surfaces in the (100) direction, where the projections of the 2b position, the Mo atoms (2c), and the S atoms (4f), are {0}, {1/3,2/3}, {1/3,2/3}, respectively. If the surface's intercept is 0, which is possible because this cleavage does not cut atoms, then the broken Wannier functions at the 1b will contribute to the OSSs. Other possible surfaces with OSSs are detailed in Ref. [Xu, Y. et al. Three-dimensional real space invariants, obstructed atomic insulators and a new catalytic principle. Submitted (2021)].

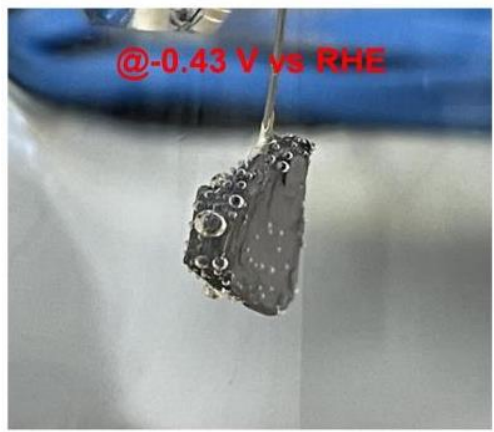

**Figure S1. a.** Photo of the 2H-MoTe$_2$ single-crystal catalyst during the HER process at an overpotential of -0.43 V *vs* RHE. The basal plane is covered by gel. It can be seen clearly that the edge surfaces are catalytically active towards HER.

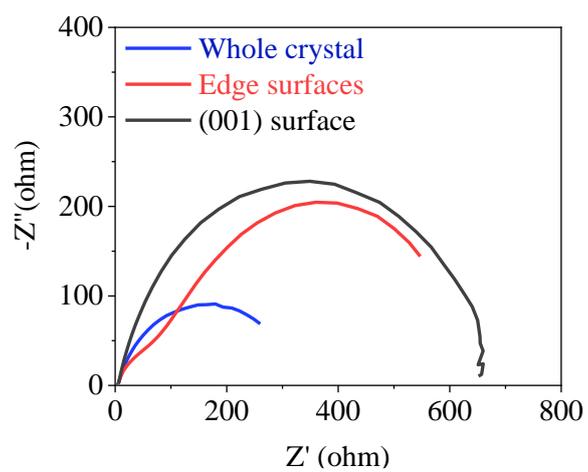

**Figure S2.** Nyquist plots of the 2H-MoTe$_2$ single-crystal catalyst during the HER process. It can be seen that the charge resistance at the edge surfaces is smaller than that for the basal plane.

| Multiplicity | Wyckoff letter | Site symmetry | Coordinates |
| --- | --- | --- | --- |
| | | | (0,0,0) + (1/2,1/2,0) + |
| 8 | j | 1 | (x,y,z) (-x,y,-z) (-x,-y,-z) (x,-y,z) |
| 4 | i | m | (x,0,z) (-x,0,-z) |
| 4 | h | 2 | (0,y,1/2) (0,-y,1/2) |
| 4 | g | 2 | (0,y,0) (0,-y,0) |
| 4 | f | -1 | (1/4,1/4,1/2) (3/4,1/4,1/2) |
| 4 | e | -1 | (1/4,1/4,0) (3/4,1/4,0) |
| 2 | d | 2/m | (0,1/2,1/2) |
| 2 | c | 2/m | (0,0,1/2) |
| 2 | b | 2/m | (0,1/2,0) |
| 2 | a | 2/m | (0,0,0) |

**Table S3.** The Wyckoff positions and their site symmetry groups of SG 12 (*C*2/*m*).

| Index | Wyckoff positions | Irreps and multiplicity | | | | | | | | | | | | mod |
|---|---|---|---|---|---|---|---|---|---|---|---|---|---|---|
| | | $\bar{\Gamma}_3\bar{\Gamma}_4$ | $\bar{\Gamma}_5\bar{\Gamma}_6$ | $\bar{A}_3\bar{A}_4$ | $\bar{A}_5\bar{A}_6$ | $\bar{L}_2\bar{L}_2$ | $\bar{L}_3\bar{L}_3$ | $\bar{M}_3\bar{M}_4$ | $\bar{M}_5\bar{M}_6$ | $\bar{V}_2\bar{V}_2$ | $\bar{V}_3\bar{V}_3$ | $\bar{Y}_3\bar{Y}_4$ | $\bar{Y}_5\bar{Y}_6$ | |
| $\delta_1$ | a | 0 | 1/4 | 0 | 1/4 | 1/2 | 0 | -1 | -3/4 | 1/2 | 0 | 0 | 1/4 | |
| $\delta_2$ | b | 0 | 1/4 | 0 | 1/4 | -1/2 | 0 | 0 | 1/4 | -1/2 | 0 | 0 | 1/4 | |
| $\delta_3$ | c | 0 | 1/4 | 0 | -1/4 | -1/2 | 0 | 0 | -1/4 | 1/2 | 0 | 0 | 1/4 | |
| $\delta_4$ | d | 0 | 1/4 | 0 | -1/4 | 1/2 | 0 | 0 | -1/4 | -1/2 | 0 | 0 | 1/4 | |
| $\delta_5$ | e | 0 | -1/4 | 0 | -1/4 | 0 | 0 | 0 | 1/4 | 0 | 0 | 0 | 1/4 | |
| $\delta_6$ | f | 0 | -1/4 | 0 | 1/4 | 0 | 0 | 0 | -1/4 | 0 | 0 | 0 | 1/4 | |
| $\eta_1$ | g,h,i | 0 | 3/2 | 0 | 1/2 | 0 | 0 | 1 | 1/2 | 0 | 0 | 0 | 3/2 | mod 2 |

**Table S4:** The formula of the RSIs of SG 12 (*C*2/*m*).

NiPS$_3$ has the symmetries of the SG 12 ($C2/m$). Ni and P atoms occupy the 4g and 4i Wyckoff positions (defined in Table S3), respectively, and the S atoms occupy 4i and 8j Wyckoff positions. The band structure and the band representation analysis of this material, which show it is a topologically trivial insulator with an indirect gap ~0.005eV and a smallest direct gap ~0.02eV, can be found on www.topologicalquantumchemistry.com. We use the symmetry-data-vector to represent the irreps formed by the occupied Bloch bands at high symmetry momenta. In the orders of the irreps listed in Table S4, the symmetry-data-vector is

$$B = (16,17,16,17,17,16,16,17,17,16,16,17)^T$$

The expressions to calculate RSIs at all Wyckoff positions are defined in table S4. Nonzero RSIs at a Wyckoff position imply Wannier functions pinned at this position. (See [Xu, Y. et al. Three-dimensional real space invariants, obstructed atomic insulators and a new catalytic principle. Submitted (2021)] for the rigorous definition of RSIs.) Applying the RSI formulae in Table S4, we find that only the 2a position has a nonzero RSI, $\delta_1 = 1$. Since the 2a position is not occupied by atoms, NiPS$_3$ is in the OAI phase.

Our theory work [Xu, Y. et al. Three-dimensional Real Space Invariants and Obstructed Atomic insulators. Submitted (2021)] shows that OSSs exist on surfaces in the directions (1-11), (111), (011), (-111), (110), (1-10), (01-1), (11-1). Here we only take the (110) direction to exemplify the analysis. We assume Ni atoms occupying the 4g position are at

$$(0, y_1, 0) \quad (0, -y_1, 0) \quad (1/2, y_1 + 1/2, 0) \quad (1/2, -y_1 + 1/2, 0)$$

P atoms occupying the 4i position are at

$$(x_2, 0, z_2) \quad (-x_2, 0, -z_2) \quad (1/2 + x_2, 1/2, z_2) \quad (1/2 - x_2, 1/2, -z_2)$$

S atoms occupying the 4i position are at

$$(x_3, 0, z_3) \quad (-x_3, 0, -z_3) \quad (1/2 + x_3, 1/2, z_3) \quad (1/2 - x_3, 1/2, -z_3)$$

S atoms occupying the 8j position are at

$$(x_4, y_4, z_4) \quad (x_4, -y_4, z_4) \quad (-x_4, y_4, -z_4) \quad (-x_4, -y_4, -z_4) \quad (1/2 + x_4, 1/2 + y_4, z_4)$$
$$(1/2 + x_4, 1/2 - y_4, z_4) \quad (1/2 - x_4, 1/2 + y_4, -z_4) \quad (1/2 - x_4, 1/2 - y_4, -z_4)$$

The projections of the atoms in the (110) direction are $\{\pm 1, \pm x_2, \pm x_3, \pm(x_4+y_4), \pm(x_4-y_4)\}$ (mod 1), while the projection of the 2a position, where the Wannier functions locate, is 1/2 (mod 1). Thus a cleavage cutting the 2a position but not the atoms is possible. On this surface, the broken Wannier functions at the 2a position will contribute to the OSSs.

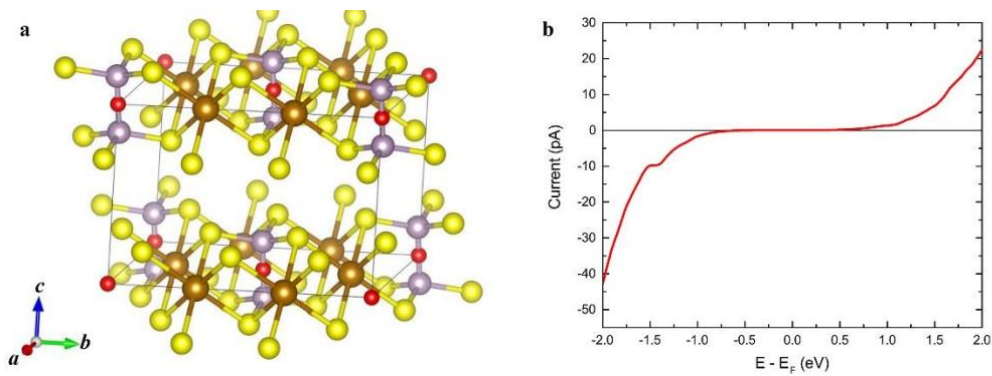

**Figure S3. a.** Crystal structure of NiPS$_3$ and the positions of OWCCs. The brown, purple, yellow, and red spheres represent the Ni, P, S atoms and the OWCCs, respectively. **b.** d$I$/d$V$ spectra taken at the surface of NiPS$_3$, indicating a bandgap of 1.6 eV.

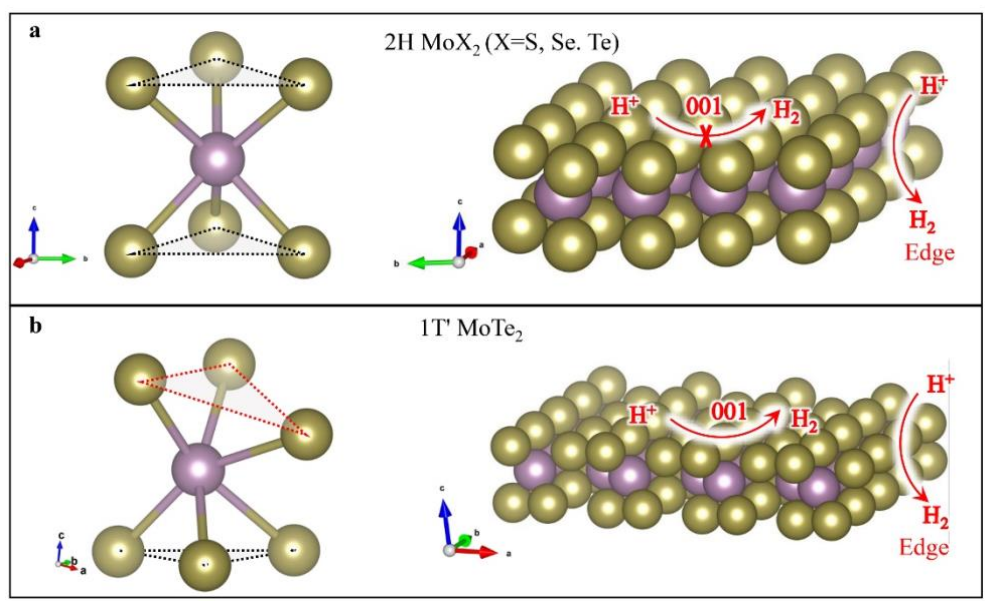

**Figure S4.** a. Crystal structure of 2H phase MoX$_2$ (X = S, Se, Te) (left) and the corresponding catalytic active surfaces (right). Mo and Te atoms are shown as purple and brown spheres, respectively. b. Crystal structure of 1T′ phase MoTe$_2$ and the corresponding catalytic active surfaces. One can see the distortion of the [MoTe$_6$] unit.

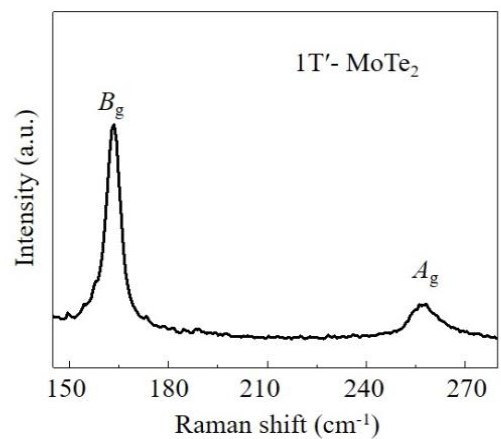

**Figure S5.** Raman spectrum of 1T′-MoTe$_2$ recorded at room temperature using a HeNe-laser (wavelength 632 nm) as the excitation source. The bands centered around 164 and 257 cm$^{-1}$ can be assigned to $B_g$ and $A_g$ modes of 1T′-MoTe$_2$, respectively

| Multiplicity | Wyckoff letter | Site symmetry | Coordinates |
|---|---|---|---|
| 4 | e | 1 | (x,y,z) (-x,y+1/2,-z+1/2) (-x,-y,-z) (x,-y+1/2,z+1/2) |
| 2 | d | -1 | (1/2,0,1/2) (1/2,1/2,0) |
| 2 | c | -1 | (0,0,1/2) (0,1/2,0) |
| 2 | b | -1 | (1/2,0,0) (1/2,1/2,1/2) |
| 2 | a | -1 | (0,0,0) (0,1/2,1/2) |

**Table S5:** The Wyckoff positions and their site symmetry groups of SG 14 ($P2_1/c$).

| Index | Wyckoff positions | Irreps and multiplicity | | | | | | | | | | | | | | |
|---|---|---|---|---|---|---|---|---|---|---|---|---|---|---|---|---|
| | | $\bar{\Gamma}_3\bar{\Gamma}_4$ | $\bar{\Gamma}_5\bar{\Gamma}_6$ | $\bar{A}_2\bar{A}_2$ | $\bar{B}_2\bar{B}_2$ | $\bar{C}_2\bar{C}_2$ | $\bar{D}_3\bar{D}_3$ | $\bar{D}_4\bar{D}_4$ | $\bar{D}_5\bar{D}_5$ | $\bar{D}_6\bar{D}_6$ | $\bar{E}_3\bar{E}_3$ | $\bar{E}_4\bar{E}_4$ | $\bar{E}_5\bar{E}_5$ | $\bar{E}_6\bar{E}_6$ | $\bar{Y}_3\bar{Y}_4$ | $\bar{Y}_5\bar{Y}_6$ | $\bar{Z}_2\bar{Z}_2$ |
| $\delta_1$ | a | 1/4 | 0 | 0 | 0 | -1 | 1/2 | 0 | 0 | 0 | 1/2 | 0 | 0 | 0 | 1/4 | 0 | 0 |
| $\delta_2$ | b | 1/4 | 0 | 0 | 0 | 0 | 1/2 | 0 | 0 | 0 | -1/2 | 0 | 0 | 0 | -1/4 | 0 | 0 |
| $\delta_3$ | c | 1/4 | 0 | 0 | 0 | 0 | -1/2 | 0 | 0 | 0 | -1/2 | 0 | 0 | 0 | 1/4 | 0 | 0 |
| $\delta_4$ | d | 1/4 | 0 | 0 | 0 | 0 | -1/2 | 0 | 0 | 0 | 1/2 | 0 | 0 | 0 | -1/4 | 0 | 0 |

**Table S6:** The formula of the RSIs of SG 14 ($P2_1/c$).

RuP$_4$ and FeP$_4$ has the symmetries of the SG 14 ($P2_1/c$). The Fe atoms occupy a 2a position and a 4e position, and the P atoms occupy six different 4e positions (defined in Table S5). The band structure and the band representation analysis of this material, which show it is a topologically trivial insulator with a gap ~ 0.9eV, can be found on www.topologicalquantumchemistry.com[6]. We use the symmetry-data-vector to represent the irreps formed by the occupied Bloch bands at high symmetry momenta. In the orders of the irreps listed in Table S6, the symmetry-data-vector is

$$B = (46,38,42,42,42,22,22,20,20,23,23,19,19,44,40,42)^T$$

The expressions to calculate RSIs at all Wyckoff positions are defined in Fig 3b. Nonzero RSIs at a Wyckoff position imply Wannier functions pinned at this position. (See [Xu, Y. et al. Three-dimensional real space invariants, obstructed atomic insulators and a new catalytic principle. Submitted (2021)] for the rigorous definition of RSIs.) Applying the RSI formulae in Table S6, we find that only the 2d position has a nonzero RSI, $\delta_4 = 1$. Since the 2d position is not occupied by atoms, RuP$_4$ and FeP$_4$ are in the OAI phase.

The ref of [Xu, Y. et al. Three-dimensional Real Space Invariants and Obstructed Atomic insulators. Submitted (2021)] shows that OSSs exist on surfaces in the directions as shown in Table S6. Here we only take the (100) direction to exemplify the analysis. We assume the 4e positions occupied by Fe are

$(x_1, y_1, z_1)$ $(-x_1 + 1/2, y_1, -z_1 + 1/2)$ $(-x_1, -y_1, -z_1)$ $(x_1 + 1/2, -y_1, z_1 + 1/2)$

and the six different 4e positions occupied by the P atoms are

$(x_i, y_i, z_i)$ $(-x_i + 1/2, y_i, -z_i + 1/2)$ $(-x_i, -y_i, -z_i)$ $(x_i + 1/2, -y_i, z_i + 1/2)$

For $i = 2 \cdots 7$. The projection of the Fe atoms at the 2a position in the (100) direction is 0, and projections of the Fe (Ru) and P atoms at the 4e positions are $\{x_i, -x_i, x_i + 1/2, -x_i + 1/2\}$ for $i = 1 \cdots 7$. While the projection of the 2d position, where the Wannier functions locate, is 1/2 (mod 1). Thus a cleavage cutting the 2d position but not the atoms is possible. On this surface, the broken Wannier functions at the 2a position will contribute to the OSSs.

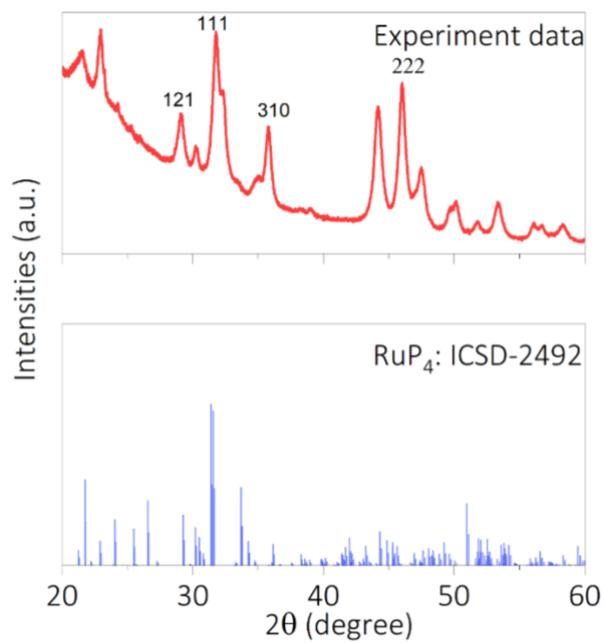

**Figure S6.** XRD pattern of the synthesized RuP$_4$ nanostructures. Up: experimental data, down: simulation data.

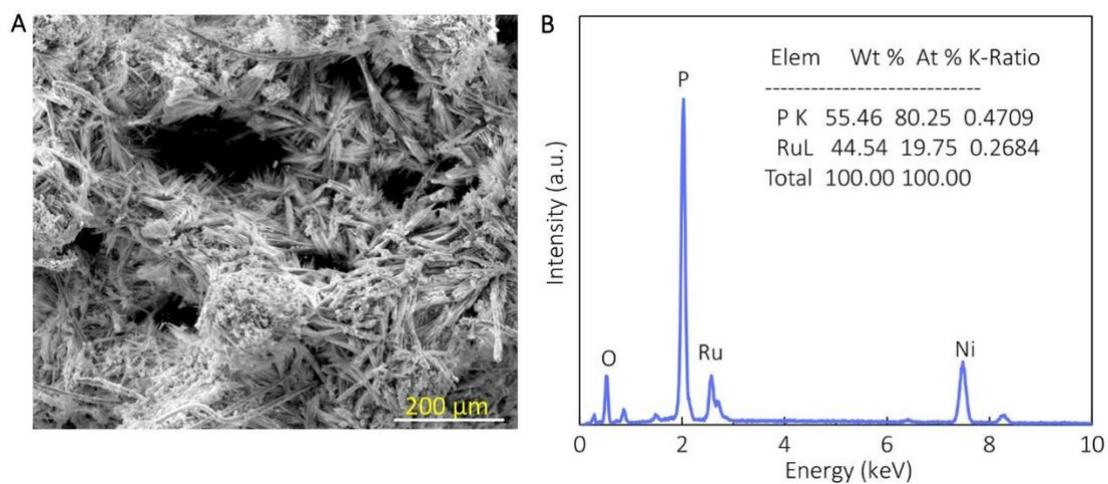

**Figure S7.** SEM (A) and EDS spectra (B) of the synthesized RuP$_4$ nanostructures. The small particle size ensured an increased exposing of surface areas, making the density of active sites as much as possible. EDS results indicate an elemental ratio of 1:4 between Ru and P, which is close to the stoichiometry of RuP$_4$.

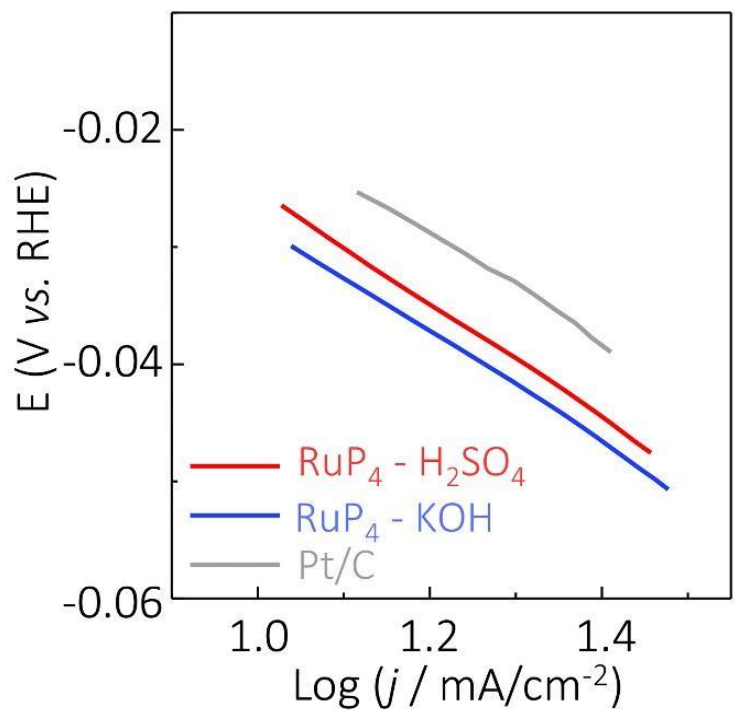

**Figure S8.** Tafel analysis reveals a similar Tafel slope of ~ 30 mV/dec between the state-of-the-art Pt/C catalyst and the OAI RuP$_4$ catalysts. This indicates that the HER takes place through the Volmer-Tafel mechanism and the recombination step is the rate-determining step for both Pt and RuP$_4$.

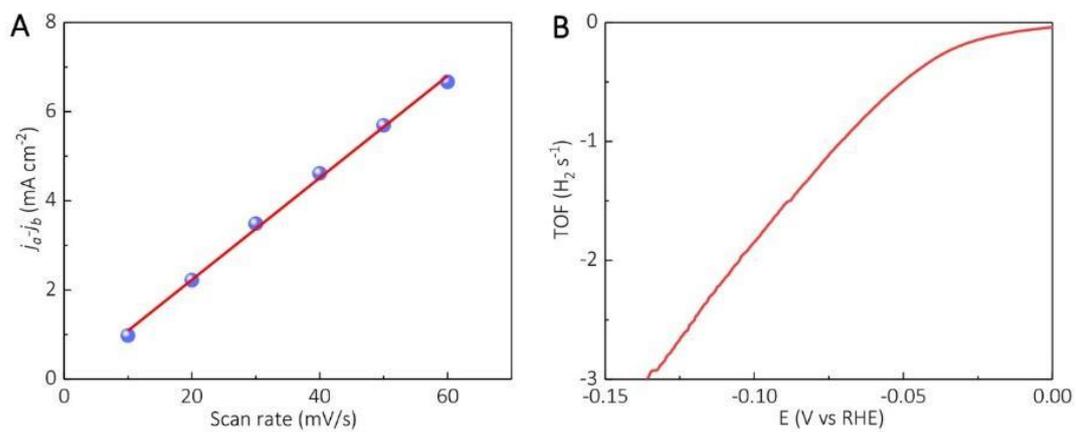

**Figure S9.** A. The double-layer capacitances ($C_{dl}$) of RuP$_4$ catalyst by measuring the capacitive currents against the scan rate in 1 M KOH solution. B. The TOF of the catalyst as a function of overpotential.

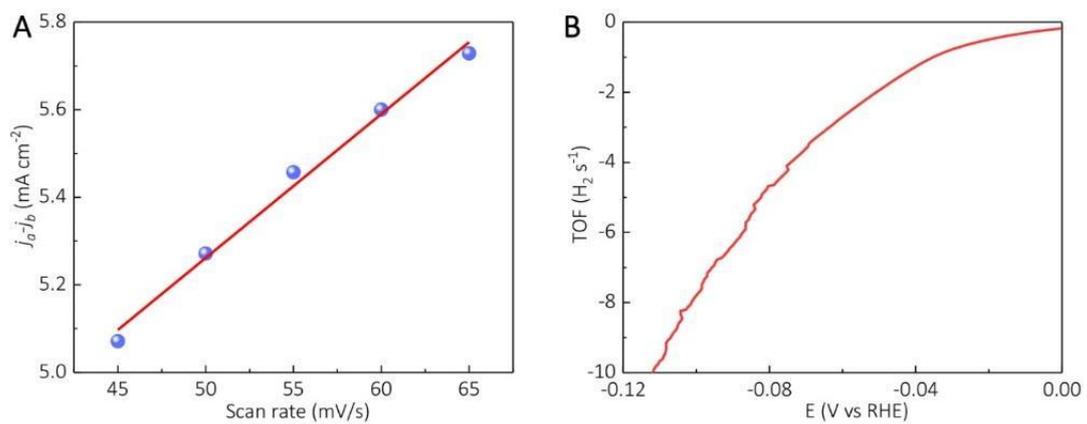

**Figure S10.** A. The double-layer capacitances (Cdl) of RuP$_4$ catalyst by measuring the capacitive currents against the scan rate in 0.5 M H$_2$SO$_4$ solution. B. The TOF of the catalyst as a function of overpotential.

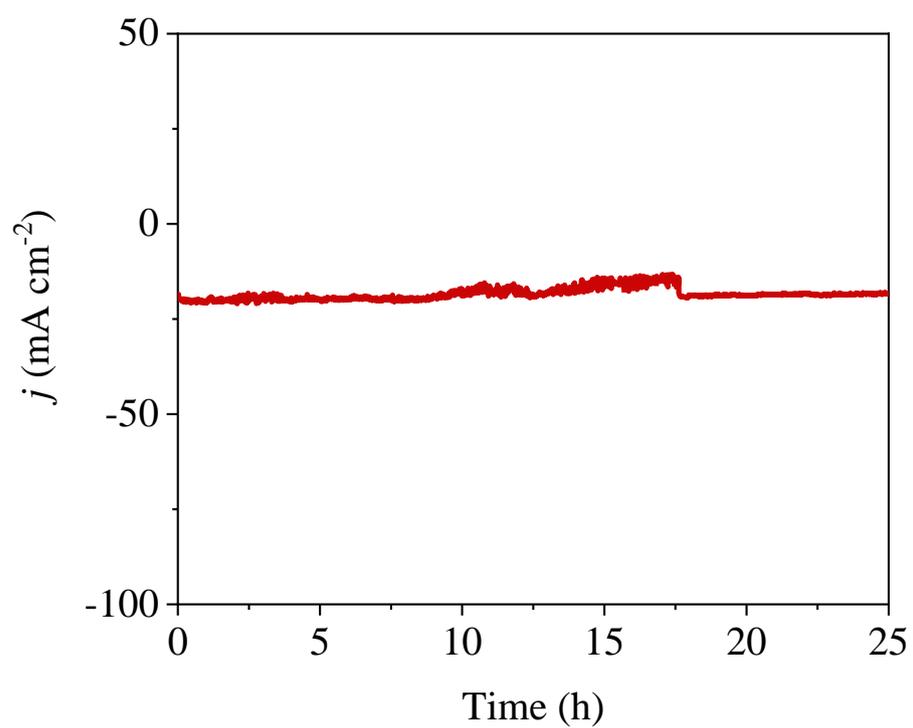

**Figure S11.** Stability test of the RuP$_4$ nanostructured catalyst at a constant overpotential of 70 mV in 0.5 M H$_2$SO$_4$ electrolyte (Without iR correction).

| Multiplicity | Wyckoff letter | Site symmetry | Coordinates |
|---|---|---|---|
| 24 | d | 1 | (x,y,z)  (-x+1/2,-y,z+1/2) (-x,y+1/2,-z+1/2) (x+1/2,-y+1/2,-z)<br>(z,x,y)  (z+1/2,-x+1/2,-y) (-z+1/2,-x,y+1/2) (-z,x+1/2,-y+1/2)<br>(y,z,x)  (-y,z+1/2,-x+1/2) (y+1/2,-z+1/2,-x) (-y+1/2,-z,x+1/2)<br>(-x,-y,-z) (x+1/2,y,-z+1/2)  (x,-y+1/2,z+1/2)  (-x+1/2,y+1/2,z)<br>(-z,-x,-y) (-z+1/2,x+1/2,y) (z+1/2,x,-y+1/2) (z,-x+1/2,y+1/2)<br>(-y,-z,-x) (y,-z+1/2,x+1/2)  (-y+1/2,z+1/2,x)  (y+1/2,z,-x+1/2) |
| 8 | c | .3. | (x,x,x)   (-x+1/2,-x,x+1/2) (-x,x+1/2,-x+1/2) (x+1/2,-x+1/2,-x)<br>(-x,-x,-x) (x+1/2,x,-x+1/2) (x,-x+1/2,x+1/2) (-x+1/2,x+1/2,x) |
| 4 | b | .-3. | (1/2,1/2,1/2) (0,1/2,0) (1/2,0,0) (0,0,1/2) |
| 4 | a | .-3. | (0,0,0) (1/2,0,1/2) (0,1/2,1/2) (1/2,1/2,0) |

**Table S7.** The Wyckoff positions and their site symmetry groups of SG 205 ($Pa\bar{3}$).

| Index | Wyckoff positions | Irreps and multiplicity | | | | | | | | | | | | |
|---|---|---|---|---|---|---|---|---|---|---|---|---|---|---|
| | | $\bar{\Gamma}_5$ | $\bar{\Gamma}_6\bar{\Gamma}_7$ | $\bar{\Gamma}_8$ | $\bar{\Gamma}_{10}\bar{\Gamma}_9$ | $\bar{M}_3\bar{M}_3$ | $\bar{M}_4\bar{M}_4$ | $\bar{R}_4\bar{R}_4$ | $\bar{R}_5\bar{R}_6$ | $\bar{R}_7\bar{R}_7$ | $\bar{R}_8\bar{R}_9$ | $\bar{R}_{10}\bar{R}_{10}$ | $\bar{R}_{11}\bar{R}_{11}$ | $\bar{X}_3\bar{X}_4$ |
| $\delta_1$ | a | 0 | 0 | -1/4 | 1/2 | 0 | 0 | -1 | 0 | 0 | 0 | 0 | 0 | 0 |
| $\delta_2$ | a | 0 | 0 | 1/2 | 0 | -1 | 0 | 1 | 0 | 1 | 1 | 0 | 0 | 0 |
| $\delta_3$ | b | 0 | 0 | -1/4 | 1/2 | 0 | 0 | 0 | 0 | -1 | 0 | 0 | 0 | 0 |
| $\delta_4$ | b | 0 | 0 | 1/2 | 0 | 0 | 0 | 0 | 0 | 0 | -1 | 0 | 0 | 0 |
| $\delta_5$ | c | 0 | 0 | -1 | 1 | 1 | 0 | -3 | 0 | -3 | 0 | 0 | 0 | 0 |

**Table S8.** The formula of the RSIs of SG 205 ($Pa\bar{3}$)

FeS$_2$ has the symmetries of the SG 205 ($Pa\bar{3}$). The Fe atoms occupy a 4a position, and the S atoms occupy an 8c position (defined in Table S7). The band structure and the band representation analysis of this material, which show it is a topologically trivial insulator with a gap ~0.6 eV, can be found on www.topologicalquantumchemistry.com. We use the symmetry-data-vector to represent the irreps formed by the occupied Bloch bands at high symmetry momenta. In the orders of the irreps listed in Table S10, the symmetry-data-vector is

$$B = (10,9,4,4,10,10,2,4,1,3,6,4,20)^T$$

The expressions to calculate RSIs at all Wyckoff positions are defined in Fig 6b. Nonzero RSIs at a Wyckoff position imply Wannier functions pinned at this position. (See [Xu, Y. et al. Three-dimensional real space invariants, obstructed atomic insulators and a new catalytic principle. Submitted (2021)] for the rigorous definition of RSIs.) Applying the RSI formulae in Table S8, we find that only the 4b position has a nonzero RSI, $\delta_3 = -1$. Since the 2b position is not occupied by atoms, FeS$_2$ is in the OAI phase.

The ref of [Xu, Y. et al. Three-dimensional real space invariants, obstructed atomic insulators and a new catalytic principle. Submitted (2021)] shows that OSSs exist on surfaces in the direction (111). The projection of the Fe atoms at the 4a position in the (111) direction is 0, and projections of the S atoms at the 8c positions are {3x,-3x,x,-x}. While the projection of the 4b position, where the Wannier functions locate, is 1/2 (mod 1). Thus a cleavage cutting the 4b position but not the atoms is possible. On this surface, the broken Wannier functions at the 4b position will contribute to the OSSs.

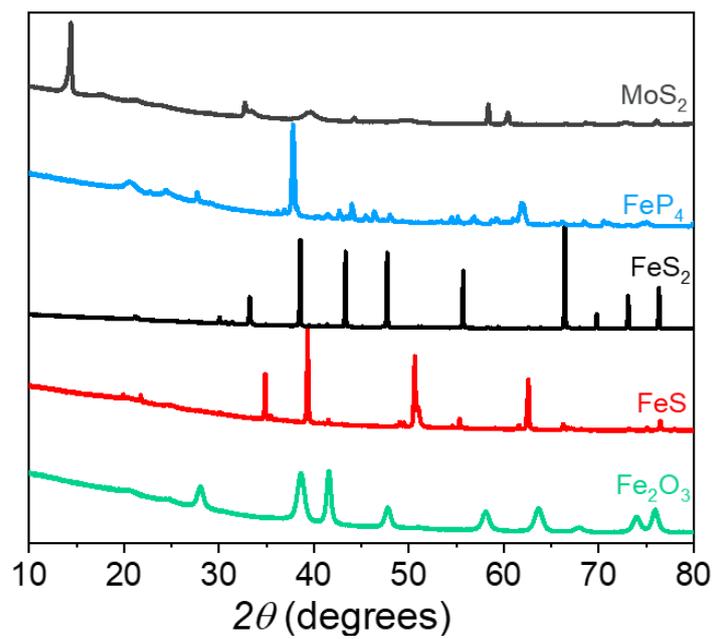

**Figure S12.** Powder X-ray diffraction patterns of the used photocatalysts, which can be well indexed to OAI $MoS_2$, $FeS_2$, $RuP_4$, and trivial insulator FeS and $Fe_2O_3$.

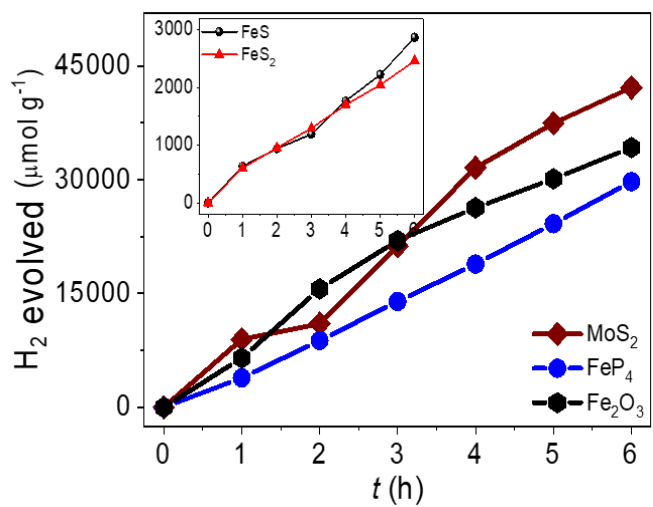

**Figure S13.** Time course plots of $H_2$ generation for obstructed insulators $FeP_4$, $MoS_2$, and $FeS_2$, and normal insulators $Fe_2O_3$ and FeS. In a hydrogen production time of 6 h, $MoS_2$ catalysts produced the largest amount of hydrogen, which is slightly higher than that of $FeP_4$ and $Fe_2O_3$. On the other hand, FeS and FeS produced much less hydrogen. Here it should be noted that we don't consider the specific areas of the catalysts.

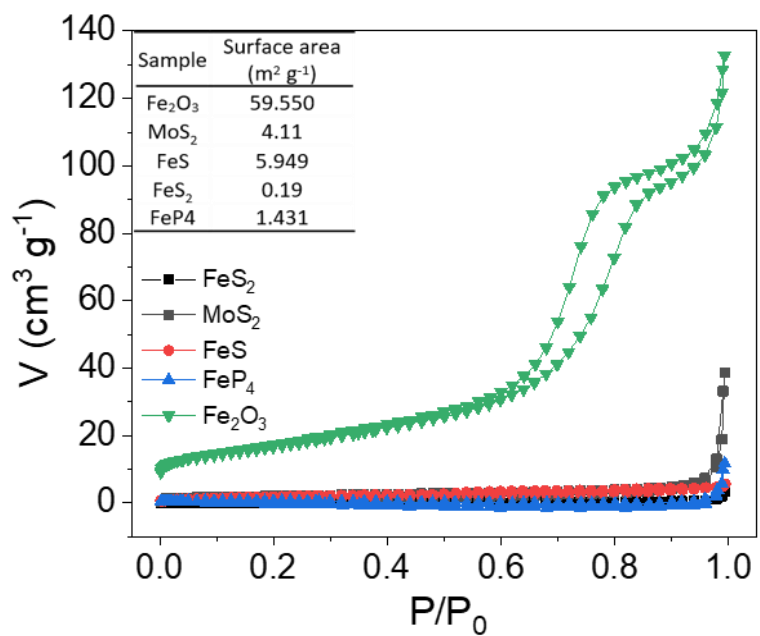

**Figure S14.** $N_2$ adsorption-desorption isotherms of the investigated catalysts and the corresponding surface areas. For a fair comparison of catalytic activities, the amounts of produced hydrogen are scaled with the surface areas of the investigated catalysts.